\documentstyle[aps,prc,preprint,epsf,tighten]{revtex}
\begin{document}
\baselineskip 14pt plus 2pt minus 2pt
\newcommand{\beq}{\begin{equation}}
\newcommand{\eeq}{\end{equation}}
\newcommand{\beqa}{\begin{eqnarray}}
\newcommand{\eeqa}{\end{eqnarray}}
\newcommand{\dfrac}{\displaystyle \frac}
\renewcommand{\thefootnote}{\#\arabic{footnote}}
\newcommand{\ve}{\varepsilon}
\newcommand{\krig}[1]{\stackrel{\circ}{#1}}
\newcommand{\barr}[1]{\not\mathrel #1}

\begin{titlepage}

%%{\bf DRAFT, \today} 

\hfill LPT 96--15

\hfill TK 96 21

%\hfill hep--ph/9507nnn

\vspace{2.0cm}

\begin{center}

{\large  \bf {
NUCLEON ELECTROWEAK FORM FACTORS:\\

\smallskip

ANALYSIS OF THEIR SPECTRAL FUNCTIONS}}

\vspace{1.2cm}
                              
{\large V. Bernard$^{\ddag}$,
N. Kaiser$^{\diamond}$, 
Ulf-G. Mei\ss ner$^{\dag}$}\footnote{Address after Oct. 1,
    1996: FZ J\"ulich, IKP (Theorie), D-52425 J\"ulich, Germany}

\vspace{1.0cm}

$^{\ddag}$Universit\'e Louis Pasteur, Laboratoire de Physique
Th\'eorique\\ BP 28, F--67037 Strasbourg, France\\
{\it email: bernard@crnhp4.in2p3.fr} \\

\vspace{0.4cm}
$^{\diamond}$Technische Universit\"at M\"unchen, Physik Department T39\\ 
James-Franck-Stra{\ss}e, D--85747 Garching, Germany\\
{\it email: : nkaiser@physik.tu-muenchen.de}\\

\vspace{0.4cm}
$^{\dag}$Universit\"at Bonn, Institut f\"ur Theoretische Kernphysik\\
 Nussallee 14-16, D--53115 Bonn, Germany\\
{\it email: meissner@itkp.uni-bonn.de} \\

\end{center}

\vspace{0.4cm}

\begin{abstract}
\noindent We investigate the imaginary parts of the nucleon
electromagnetic and axial form factors close to threshold in the
framework of heavy baryon chiral perturbation theory. For the isovector 
electromagnetic form factors, we recover the well known strong enhancement near
threshold. For the isoscalar ones, we show that there is no
visible enhancement due to the three--pion continuum. This justifies
the use of vector meson poles only in dispersion--theoretical
calculations. We also calculate the imaginary part of the nucleon isovector 
axial form factor and show that it is small in the threshold region. 
\end{abstract}

%\vspace{3.5cm}

\vspace{2cm}

\vfill

\end{titlepage}

\section{Introduction and summary}
\label{sec:intro}
The electromagnetic and axial structure of the nucleon as revealed
e.g. in elastic electron--nucleon  and (anti)neutrino--nucleon scattering
is parameterized in terms of the six
form factors $F_{1,2}^{p,n} (t)$ and $G_{A,P} (t)$\footnote{In what
follows, we will not be concerned with the induced pseudoscalar form
factor $G_P(t)$ which is dominated by its pion--pole term $4g_{\pi N} m
F_\pi/(M_\pi^2 -t)$.} (with $t$ the squared momentum transfer). The 
understanding of these form factors is of utmost importance in any theory or 
model of the strong interactions. Abundant data on these form factors over a 
large range of momentum transfer exist and this data base will increase when 
the recent/upcoming experiments performed at ELSA, MAMI and TJNAL will
have been analyzed.  Dispersion theory is a tool to interpret (and cross check)
these data in a largely  model--independent fashion.
The form factors can be written in terms of unsubtracted dispersion
relations and their absorptive parts are often parameterized in terms of  
a few vector meson poles. This procedure is based on the
successful vector meson dominance (VMD) hypothesis, which states 
that a photon (or $W/Z$--boson) couples to hadrons only
via  intermediate vector mesons. However, as already pointed out in
1959 by Frazer and Fulco \cite{frfu}, such an approach
is not in conformity with general constraints from unitarity and
analyticity. In particular the singularity structure of the triangle 
diagram is not respected at all when using only vector meson poles. As a
consequence of this singularity structure the two--pion continuum
has a pronounced effect on the isovector spectral functions on the
left wing of the $\rho$--resonance. This becomes  particularly visible in the 
determinations of the corresponding nucleon mean square 
radii. This effect  was  quantified by the Karlsruhe--Helsinki group
in their seminal work on the nucleon electromagnetic  form factors
\cite{hopi,hoeh76}. This analysis was recently refined by the
Mainz--Bonn group accounting for new data and the high energy constraints from
perturbative QCD \cite{MMD,HMD}. However, there still remains one open end 
concerning the dispersion--theoretical analysis. In the isoscalar
electromagnetic channel, it is believed (but not proven) that  the pertinent
spectral functions rise smoothly from the three--pion threshold  to the
$\omega$-meson peak, i.e. that there is no pronounced effect from the
three--pion cut on the left wing of the $\omega$-resonance (which also has a
much smaller width than the $\rho$-meson). Chiral perturbation theory 
\cite{gss,bkmrev} can be used to settle this issue. An investigation of the
isoscalar spectral functions based on pion scattering data and dispersion 
theory as done for the isovector  spectral functions \cite{hopi,hoeh76}
seems not to be feasible at the moment since it requires the full
dispersion--theoretical analysis of the three-body processes $\pi N  \to \pi\pi
N$ (or of the data on $\bar N N \to 3\pi$).  
 
The one loop calculation of the isovector nucleon form factors indeed shows the
strong unitarity correction on the left wing of the $\rho$-meson (i.e. slightly
above threshold) \cite{gss,bkmrev}. For the isoscalar form factors, a
calculation of the imaginary parts of certain two--loop diagrams will reveal 
whether there is  some enhancement above $t = 9 M_\pi^2$ or justify the common
assumption that one has a  smooth isoscalar spectral functions driven by the
$\omega$-meson at low $t$.  Similarly, the spectral function of the isovector
axial form factor is supposedly dominated by correlated $\pi\rho$-exchange (or
the $a_1$-meson) in the region above threshold. Nothing is yet known about a 
possible strong enhancement of the pertinent imaginary part slightly above 
threshold.  It is exactly this problem we wish to address in this paper by
evaluating the imaginary parts of the pertinent two loop diagrams at chiral
order $q^7$ and $q^5$ contributing to the isoscalar electromagnetic and the
isovector axial form factors in the framework of heavy baryon chiral
perturbation theory.  

\newpage

The pertinent results of this investigation can be summarized as follows:

\begin{enumerate}

\item[(i)] We have calculated the imaginary parts of the isovector
  electromagnetic form factors in heavy baryon CHPT to order $q^4$ (in the 
 effective Lagrangian), thus  extending previous studies \cite{gss,bkmrev}. 
The strong enhancement slightly above threshold on the left
 wing of the $\rho$-resonance  due to the two--pion continuum is recovered and
 good agreement with the empirical analysis is found if one subtracts from it
 the $\rho$--meson contribution (by setting the pion charge form factor
 $F_\pi^V (t) \equiv 1$) as described in section~\ref{sec:imffV}).

\item[(ii)] We have determined the leading contribution to the imaginary 
parts of the isoscalar electromagnetic form factors in the threshold region
by evaluating the imaginary part of the relevant two--loop diagrams at order 
$q^7$. As previously anticipated, these imaginary parts are very small and rise
smoothly with increasing $t$. The common procedure \cite{hoeh76} \cite{MMD} of 
describing the isoscalar electromagnetic spectral functions by the
$\omega$-meson pole (and higher mass poles and/or continua) is therefore
justified. 

\item[(iii)] We have also calculated the imaginary part of the nucleon 
isovector axial form factor in the threshold region. It shows a behavior
similar to the isoscalar electromagnetic spectral functions, i.e. a smooth 
rise from threshold. Furthermore, this three--pion contribution is numerically 
small.

\end{enumerate}

\section{Basic concepts}
\label{sec:con}
In this section, we assemble all necessary definitions for the calculation
of the imaginary parts (spectral distributions) discussed below. 
This section serves mainly the purpose of fixing the notation
underlying our analysis.
This material is mostly not new but necessary to keep the
manuscript self--contained. The reader familiar with it is invited to
skip this section.

Consider first the nucleon electromagnetic  form factors. They are
defined by the nucleon matrix element of the quark electromagnetic current,
\begin{equation} 
\langle N(p')|\bar q \gamma^\mu Q q|N(p) \rangle = \bar u(p')
\biggl[\gamma^\mu\, F_1(t) +{i\over 2m } \sigma^{\mu\nu}(p'-p)_\nu\,F_2(t)
\biggr] u(p)  
\label{defemff}
\end{equation}
with $t=(p'-p)^2$ the invariant momentum transfer squared and
$Q$ the quark charge matrix,  
$Q= {\rm diag}({2\over3},-{1\over3},-{1\over3})$. $F_1 (t) $ and
$F_2(t)$ are called the Dirac and Pauli form factors, in order. 
Following the conventions of \cite{MMD}, the electromagnetic form factors are
decomposed into isoscalar ($S$) and isovector ($V$) components
\begin{equation}
 F_{1,2}(t) = F_{1,2}^S(t) + \tau^3 F^V_{1,2}(t) \quad ,
\label{isodec}
\end{equation} 
subject to the normalization
\begin{equation} 
F_1^S(0) = F_1^V(0) = {1\over 2} \,, \quad
F_2^S(0)={\kappa_p+\kappa_n\over 2}\,, 
\quad F_2^V(0)={\kappa_p-\kappa_n\over 2} \,\,\, ,
\label{emnorm}
\end{equation}
with $\kappa_p = 1.793 \,\, (\kappa_n =-1.913)$ the anomalous magnetic
moment of the proton (neutron). In what follows, we will  work with the
electric and magnetic (Sachs) form factors,
\begin{equation}
 G^{S,V}_E(t) = F_1^{S,V}(t) + {t\over 4m^2} F_2^{S,V}(t)\,,
\quad G^{S,V}_M(t) = F_1^{S,V}(t)+F_2^{S,V}(t) 
\label{sachs}
\end{equation}
in the isospin basis, and $m=938.27$ MeV denotes the nucleon mass. 

Similarly, the nucleon matrix element of the quark isovector axial
current is parameterized in terms of two form factors,
\begin{equation} 
\langle N(p')|\bar q \gamma^\mu \gamma_5 \tau^a q|N(p) \rangle
= \bar u(p') \biggl[\gamma^\mu\, G_A(t) +{1\over 2m }(p'-p)^\mu\,G_P(t) \biggr]
\gamma_5 \tau^a u(p)  \,\, ,
\label{defax}
\end{equation}
with $G_A(t)$ called the axial and $G_P(t)$ the induced pseudoscalar
form factor. In what follows, we will only be concerned with the axial
form factor. Its normalization  is given by $G_A(0)=g_A=1.26$.

We briefly discuss now the pertinent dispersion relations.
Let $F(t)$ be a generic symbol for any one of the five nucleon
electroweak form factors. We assume the validity of an unsubtracted
dispersion relation of the form
\beq
F(t) = \dfrac{1}{\pi} \, \int_{t_0}^\infty \dfrac{{\rm Im} \,
  F(t')}{t'-t - i\epsilon} \, dt' \, \, \, , 
\label{disp} \eeq
where  the $-i \epsilon$ in the numerator is necessary if $t>t_0$, since in
that region  $F(t)$ is complex valued.
The spectral function Im~$F(t)$ is different from zero along the
cut from $t_0$ to $+\infty$ on the real $t$-axis, with $t_0 = 9  \, M_\pi^2$
for the isoscalar electromagnetic and the axial case. $M_\pi= 139.57$ MeV
denotes the pion mass and we neglect isospin breaking effects in the pion and
nucleon masses. For the isovector electromagnetic form factors the
threshold lies at $t_0= 4 \, M_\pi^2$. The
proof of the validity of such dispersion relations in QCD has not yet
been given \cite{oehme}. Eq.(\ref{disp}) means that the electroweak structure
of the nucleon is entirely determined from its absorptive behavior,
with the data for $F(t)$  given (mainly) for $t\leq 0$. In what follows, we 
will concentrate on the threshold behavior of the various spectral
functions, i.e. the range $ t_0 \le t \le 30\,M_\pi^2$. The upper end
of this range corresponds to the mass of the first isoscalar and
isovector vector mesons, i.e. the $\omega$- and the $\rho$-meson 
($M_\omega^2 = 31.4\, M_\pi^2$, \, \,$M_\rho^2 = 30.3 \, M_\pi^2$).  

\section{Isovector electromagnetic spectral functions}
\label{sec:imffV}

Unitarity determines the spectral functions of the isovector electromagnetic 
form factors in the interval $4M_\pi^2 <t<16 M_\pi^2$ uniquely as,\footnote{In
practise, this representation  holds up to $t \simeq 50\,M_\pi^2$.}
\begin{eqnarray} 
{\rm Im}\, G_E^V(t) &=&  {(t-4M_\pi^2)^{3/2}\over 8m \sqrt t} \,
f^1_+(t)\, {F_\pi^V}^* (t)\,, \nonumber \\ 
{\rm Im}\,G_M^V(t) &=& {(t-4M_\pi^2)^{3/2}\over 8
\sqrt{2 t}} \, f^1_-(t)\, {F_\pi^V}^* (t) \,\, ,
\label{unit}
\end{eqnarray}
with $F_\pi^V (t)$ the pion charge form factor and the $f_\pm^1 (t)$ are the 
P--wave $\pi N$ partial wave amplitudes in the $t$--channel ($\pi\pi \to 
\bar N N$), cf. fig.~1. The partial wave amplitudes $f_\pm^1 (t)$ have a 
logarithmic singularity on the second Riemann sheet \cite{hoeh76} (originating
from the projection of the nucleon pole terms in the invariant $\pi N$
scattering amplitudes) located at 
\beq
t_c = 4 M_\pi^2 -M_\pi^4 / m^2 = 3.98 \, M_\pi^2 \, \, ,
\eeq
very close to the physical threshold at $t_0 = 4 M_\pi^2$. The isovector form 
factors inherit this  singularity (on the second sheet) and the closeness to
the physical threshold leads to the pronounced enhancement of the isovector 
spectral functions weighted with $1/t^2$ as shown in fig.~2. Note that the 
ratio of the magnetic to the electric spectral function is very close to 
$1+\kappa_p-\kappa_n = 4.71$, i.e. the ratio of the form factors in dipole 
approximation. For later comparison, we also show the corresponding curves with
$F_\pi^V (t) \equiv 1$ (to suppress the contribution from the $\rho$--meson). 
This strong enhancement is obviously very important for a precise determination
of the nucleon isovector mean square radii \cite{hopi,MMD}.

In the framework of relativistic baryon CHPT, the isovector spectral
functions were calculated in \cite{gss}. The correct analytic
structure naturally emerged in this calculation and thus the strong
threshold enhancement. However, in that framework there is no consistent power
counting scheme due to the extra energy scale related to the nucleon mass
in the chiral limit. This can be overcome in the heavy baryon
approach, which is discussed in detail in the review \cite{bkmrev}. In
that paper, the imaginary parts of the isovector Dirac and Pauli form factors
were calculated to order $q^3$ and compared to the result of \cite{gss}. Here, 
we go one step further and calculate also the first corrections to these 
results at order $q^4$, thus completing the one loop calculation. The 
imaginary part of the isovector electric form factor can be given entirely in 
terms of lowest order parameters 
\begin{equation} 
{\rm Im}\, G^V_E(t) = {\sqrt{t-4M_\pi^2}\over 192 \pi
F_\pi^2\sqrt t} \biggr[ t - 4M_\pi^2 + g_A^2(5t-8M_\pi^2)\biggr]
-{g_A^2(t-2M_\pi^2)^2 \over 128mF_\pi^2 \sqrt t} \,\,\, ,
\label{imgev}
\end{equation} 
where the term in the square brackets is the order $q^3$ result and
$F_\pi = 92.4\,$ MeV is the pion decay constant.
We remark that in
the expressions for the imaginary parts we 
make use of  the Goldberger-Treiman relation to determine the value of
$g_A$, $g_A= g_{\pi N}
F_\pi/m=1.32$, since here the pion-nucleon coupling constant $g_{\pi N}=13.4$ 
appears naturally and not the axial current coupling $g_A$.  
The expression for the imaginary part of the isovector magnetic form factor to
order $q^4$ contains the low--energy  constant $c_4$ of the dimension two
pion--nucleon Lagrangian (we take the value $c_4 = 2.25\,$GeV$^{-1}$ determined
in \cite{bkmpipin} from some $\pi N$ data),
\begin{equation} 
{\rm Im}\,G^V_M(t) = {g_A^2 m(t - 4M_\pi^2) \over 64 F_\pi^2 
\sqrt t  } + {\sqrt{t-4M_\pi^2}\over 192 \pi F_\pi^2 \sqrt t} \biggr[
(1+4mc_4)(t - 4M_\pi^2) + g_A^2(16M_\pi^2-7t)\biggr]  \,\, .
\end{equation}   
We remark that at this order the normal threshold $t_0=4M_\pi^2$ and the 
anomalous threshold $t_c = 4M_\pi^2-M_\pi^4/m^2$ on the second Riemann sheet 
still coalesce (it requires accuracy $q^5$ to separate $t_0$ and $t_c$). As a
consequence of this, the isovector imaginary parts show an abnormal threshold
behavior in the $1/m$ expansion, i.e. they do not start out as $(t-4M_\pi^2
)^{3/2}$ as in eq.(7). The coalescence of both thresholds leads to smaller
exponents. This problem is, however, largely academic as the
comparison of the so calculated imaginary parts (weighted with
$1/t^2$) shown in fig.~3 with the dashed--dotted lines of fig.~2
reveals (one should only compare to the empirical spectral
functions with the contribution of the $\rho$-meson subtracted since this
resonance is not included in the effective field theory). In fig.~3, the
results of the order $q^3$ and order $q^4$ calculations are
shown separately by the dashed and solid lines, respectively,
indicating that the empirical observation of the strong peak close to
threshold  can indeed be explained within the heavy mass 
expansion of the chiral 
effective pion--nucleon field theory. This shows that the method can be
used to investigate the isoscalar spectral distributions in the
threshold region (below the $\omega$-resonance).

\section{Isoscalar electromagnetic spectral functions}
\label{sec:imffS}

The imaginary parts of the isoscalar electromagnetic form factors open at the
three--pion threshold $t_0 = 9\,M_\pi^2$. The three--pion cut contribution is 
depicted in fig.~4. In the $\bar N N$ center-of-mass system, the
pertinent anti-nucleon and nucleon four--momenta are
$p_1^\mu=(\sqrt{t}/2, \vec p \,)$ and $p_2^\mu=(\sqrt{t}/2,-\vec p \,)$.
The analytic continuation of the three--momentum $\vec p$ into the region $t
<4\, m^2$ takes the form
\beq
\vec p \, = \hat p \, \sqrt{t/4 -m^2} = i\, m \, \hat p 
+ \ldots \,\, .
\eeq
in the heavy mass expansion, with $\hat p$ a real unit vector. Application of 
the  Cutkosky rules to three--pion intermediate state with four--momenta $l_1$,
$l_2$ and $l_3=p_1+p_2-l_1-l_2$ (cf. fig.~4) leads to 
\begin{equation} 
{\rm Im}\,\biggl[ A \,\int{d^4l_1\over i(2\pi)^4}
\int{d^4l_2\over i(2\pi)^4} {1\over M_\pi^2-l_1^2} {1\over M_\pi^2-l_2^2}{1
\over M_\pi^2-l_3^2}\,B\biggr]={1\over 2}\int d\Gamma_3(A\, B) \,\, ,
\label{cut}
\end{equation} 
where the symbol 'A' refers to the $\gamma \to 3\pi$ and 'B' to the $3\pi
\to  \bar N N$ transition, respectively, and $d\Gamma_3$ is the measure on
the invariant three--body phase space. In what follows, we have to evaluate
integrals over the three--pion phase space. Tensorial integrals can be
reduced to scalar ones by the following formula  
(or any permutation of it in the indices '1' and '2')
\beqa
\int d\Gamma_3 \, H(\ldots) \, \hat{l}_1^i \, \hat{l}_1^j \,
\hat{l}_2^k &=& \int d\Gamma_3 \, H(\ldots) \, \biggl[ \frac{y}{2}
(1-x^2) \delta^{ij} \, \hat{p}^k + \frac{x}{2}(z-xy)
(\delta^{ik} \, \hat{p}^j + \delta^{jk} \, \hat{p}^i \, )
\nonumber \\ & & \qquad \qquad \qquad + 
(\frac{y}{2}(5x^2-1) -xz) \, \hat{p}^i \, \hat{p}^j \, \hat{p}^k \, \biggr]
\label{int1} \eeqa
with 
\beqa 
|\vec{l}_i| &=& \sqrt{\omega_i^2-M_\pi^2}\,\,, \, \, (i=1,2)
\,\, , \, \quad x=\hat{l}_1 \cdot \hat p 
\,\, , \, \quad y=\hat{l}_2 \cdot \hat p \,\, , \, \,  \nonumber \\ 
 z &=&\hat{l}_1 \cdot \hat{l}_2=
 \frac{\omega_1 \omega_2 -\sqrt t (\omega_1+\omega_2)
+{1\over2} (t+M_\pi^2)}{|\vec{l}_1|\,|\vec{l}_2|}\, \,\,\, ,
\eeqa
and the function $H(\ldots)$ depends on the variables
$\omega_1, \omega_2,x,y,z,t$. The integration over the three-pion phase space
can be expressed as  a four-dimensional integral of the form  
\beq
\frac{1}{2} \, \int d\Gamma_3 \, H(\ldots) \, = \frac{1}{128\pi^4}
\int\int_{z^2<1} d\omega_1 d\omega_2 \, \int \int_E 
\frac{dx dy}{\sqrt{1-x^2-y^2-z^2+2xyz}} \, H(\ldots) \, \, \, ,
\label{int2}
\eeq
where the interior of the ellipse $E$ is defined by the equation
$1-x^2-y^2-z^2+2xyz > 0 $. The ellipse has semi--axes of 
length $\sqrt{1-z}$ and $\sqrt{1+z}$, respectively. 
 
In chiral perturbation theory, to leading order $q^7$, the two--loop
diagrams shown in fig.~5 can contribute to the isoscalar imaginary parts, but
graph (d) vanishes  because of an isospin factor zero.
All (relativistic) Feynman rules needed to calculate these graphs can be
recovered from appendix~A of \cite{bkmrev} with the exception of the anomalous
$\gamma 3\pi$ vertex. In the chiral effective Lagrangian approach, it follows
from the Wess--Zumino--Witten term (for a review, see \cite{hans}),
\beq
{\cal L}^{(4)}_{\pi\pi} = \frac{e\,\epsilon^{\mu \nu \alpha \beta}}{48\pi^2} \,
A_\mu\, {\rm Tr} \, (\partial_\nu U U^\dagger\, \partial_\alpha U U^\dagger\,
\partial_\beta U U^\dagger\,) \,\,\, ,
\eeq
with $A_\mu$ the photon field and $U= 1+ i \vec \tau \cdot \vec \pi /F+
\dots$  collects the pions. Note that this vertex has dimension four
and therefore the two--loop graphs for the imaginary
parts start only at order $q^7$.
The corresponding Feynman insertion (with the
sign convention $\epsilon^{0123}=-1$) is 
\beq
\gamma3\pi-{\rm vertex:} \quad  -\frac{e\,\sqrt{t}}{4\pi^2F_\pi^3} \,
\vec{\epsilon} \cdot ( \vec{l}_1 \times \vec{l}_2 \, ) \,
\epsilon^{abc}
\eeq
where $'a,b,c'$ are pion isospin indices and $\vec \epsilon$ is the photon
polarization vector. With these rules at hand and using eqs.(11--15) one
arrives after straightforward but tedious calculations at the 
isoscalar imaginary parts in the limit $m\to \infty$. We give only the
results in this limit since these represent the genuine leading order
contributions with all higher order effects (starting at order $q^8$)
switched off. Furthermore, only in this limit the angular integrations
in eq.(\ref{int2}) can be performed analytically. We find:
\begin{equation} 
{\rm Im}\, G_E^S(t) = {3g_A^3 \,t\over (4 \pi)^5 F_\pi^6}
\int\int_{z^2<1} d \omega_1 d\omega_2\, |\vec{l}_1| \, |\vec{l}_2| \,  
\sqrt{1-z^2} \, \arccos(-z) \,\,,
\label{imges}
\end{equation}
\begin{eqnarray} 
&&{\rm Im}\, G_M^S(t) = {g_Am\over (8 \pi)^4 F_\pi^6}
\biggl\{L(t) \biggl[ 3t^2 -10 tM_\pi^2 + 2M_\pi^4 + g_A^2
\bigl(3t^2 -2 tM_\pi^2 - 2M_\pi^4 \bigr)\biggr]\nonumber\\ && \quad+W(t)\biggl[
t^3+2t^{5/2}M_\pi-39t^2 M_\pi^2 -12 t^{3/2}M_\pi^3 +65 t M_\pi^4 -50 
\sqrt t M_\pi^5 -27 M_\pi^6 \nonumber \\ && \quad +g_A^2 \bigl( 5t^3+10t^
{5/2}M_\pi- 147t^2 M_\pi^2 +36 t^{3/2}M_\pi^3 +277 t M_\pi^4 -58 \sqrt
t M_\pi^5 -135 M_\pi^6 \bigr) \biggr] \biggr\}\,\,,
\label{imgms} 
\end{eqnarray}   
with
\beqa 
L(t) &=& {M_\pi^4 \over 2t^{3/2}} \ln {\sqrt
t-M_\pi+\sqrt{t -2\sqrt t M_\pi-3M^2_\pi}\over2M_\pi}\,\, , \\
W(t) &=& {\sqrt t- M_\pi\over 96 t^{3/2}}\sqrt{t-2\sqrt t M_\pi-
3M^2_\pi} \,\,\, . 
\eeqa
Note that in the infinite nucleon mass limit Im\,$G_E^S(t)$ comes solely from
graph (c) in fig.~5  and quite astonishingly one can evaluate all integrals in
closed form for Im\,$G_M^S(t)$. For the sake of completeness we
give also the isoscalar electromagnetic spectral functions in the chiral limit
$M_\pi=0$. The resulting expressions can be given in closed form,
\begin{equation} 
{\rm Im}\, \krig G_E^S (t) = {1\over 105} \biggl({\krig{g}_A\, t \over
16 \pi F^2} \biggr)^3 \,\,, \qquad \qquad {\rm Im}\, \krig G_M^S(t) =
{\krig{g}_A(1+5\krig{g}_A^2) \krig m \, t^{5/2} \over 
6 (16 \pi)^4 F^6}\,\, , 
\label{cl}
\end{equation}
and exhibit a simple power like dependence on the variable $t$. In 
eq.(\ref{cl}) all quantities are to be taken at their values in 
the chiral limit, as denoted by the '$\krig{}$', with the exception of
the pion decay constant whose chiral limit value is called $F$.
The behavior near threshold $t_0 = 9\, M_\pi^2$ of the imaginary parts
for finite pion mass, eqs.(\ref{imges},\ref{imgms}), is 
\begin{equation} 
{\rm Im}\, G^S_E(t) \sim (\sqrt t-3M_\pi)^3\,, \qquad {\rm
Im}\, G^S_M(t) \sim (\sqrt t-3M_\pi)^{5/2} 
\end{equation}
which corresponds to a stronger growth than pure phase space  
\beq
 \int\int_{z^2<1} d\omega_1 d\omega_2 \,
|\vec{l}_1 \times \vec{l}_2|^2 \, \sim (\sqrt t-3M_\pi)^4 \,\, \, .
\eeq
This feature indicates (as in the isovector case) that in the heavy nucleon
mass limit $m \to \infty$ normal and anomalous thresholds coincide. In order to
find these singularities for finite nucleon mass $m$ an investigation of Landau
equations is necessary \cite{polk}.  By using standard techniques
\cite{polk} we are able find (at least) one anomalous threshold of diagrams (a)
and (b) at  
\begin{equation} 
\sqrt{t_c} = M_\pi \biggl(\sqrt{4-M_\pi^2/m^2}+
\sqrt{1-M_\pi^2/m^2} \biggr)\,\,, \quad \qquad t_c = 8.90\, M_\pi^2
\label{anthr}
\end{equation} 
which is very near to the (normal) threshold $t_0=9\, M_\pi^2$ and indeed
coalesces with $t_0$ in the infinite nucleon mass limit. We note that diagram
(d) does not possess this anomalous threshold $t_c=8.90\,M_\pi^2$, but only
the normal one. We do not want to go here deeper into the rather complicated
analysis of the full singularity structure of all two-loop diagrams in fig.~5
but are mainly interested in the magnitude of the
isoscalar electromagnetic imaginary parts. 
The resulting spectral distributions again weighted with $1/t^2$ are
shown in fig.~6. They show a smooth rise and are two orders of magnitude
smaller than the corresponding isovector ones, cf. fig.~3. This
smallness justifies the procedure in the dispersion--theoretical
analysis like in \cite{MMD} to describe the isoscalar spectral
functions solely by vector meson poles starting with the $\omega$-meson 
in the low energy region. Nevertheless, it may be worthwhile to include these
calculated isoscalar imaginary parts in future dispersion analyses.  
We finally remark, that Im~$G_{E,M}^S(t)/t^4$
which have the same asymptotic behavior (for $t\to \infty$) as
Im~$G^V_{E,M}(t)/t^2$ (considering only the leading $q^3$ contribution) do
still not show any strong peak below the $\omega$--resonance.  
Im~$G^S_E(t)/t^4$
is monotonically increasing from $t_0= 9\,M_\pi^2$ to $t=30\,M_\pi^2$ and
Im~$G_M^S(t)/t^4$ develops some plateau between $t= 20$ and $30 \,M_\pi^2$.    
This observation is a further indication  that there is indeed no enhancement 
of the isoscalar electromagnetic spectral function near threshold. 
Even though the isoscalar and isovector
electromagnetic form factors behave formally very similar concerning the 
existence of anomalous thresholds $t_c$ very close to the normal thresholds
$t_0$, the influence of these on the physical spectral functions is
rather different for the two cases. Only in the isovector case a strong
enhancement is visible. This is presumably due to the different phase space
factors, which are $(t-t_0)^{3/2}$ and $(t-t_0)^4$ for the isovector
and isoscalar case, respectively. In latter case,  the
anomalous threshold at $t_c = 8.9\,M_\pi^2$ is thus effectively masked. 

%\newpage

\section{Isovector axial  spectral function}
\label{sec:imffA}

The calculation of the imaginary part of the axial form factor Im~$G_A
(t)$ proceeds along the same lines as outlined in the previous
section. One has to consider the same graphs shown in fig.~5 (with the
wiggly line denoting now the external axial source, i.e. the $W$--boson). 
All four graphs contribute
and performing the necessary integrations, the imaginary part 
of $G_A(t)$ in the heavy mass limit reads
\begin{eqnarray} 
&&{\rm Im}\, G_A(t) = {g_A\over 192 \pi^3 F_\pi^4} 
\int\int_{z^2<1} d\omega_1 d\omega_2 \biggl\{6g_A^2(\sqrt t\,\omega_1 -M_\pi^2)
\biggl({|\vec{l}_2| \over |\vec{l}_1|} 
+z\biggr) {\arccos(-z)\over \sqrt{1-z^2}} \nonumber \\
&& \hskip 5cm + 2g_A^2(M_\pi^2 -\sqrt t\, \omega_1-\omega_1^2) +
M_\pi^2-\sqrt t\, \omega_1+2\omega_1^2  \biggr\} \,\, . 
\label{imga}
\end{eqnarray} 
In the chiral limit $M_\pi=0$ the isovector axial spectral function shows a
simple $t^2$ dependence of the form
\begin{equation} 
{\rm Im}\,\krig G_A(t) = {\krig g_A \, t^2 \over (16\pi)^3 F^4
} \biggl[ \krig g_A^2 \biggl( K - {22\over 9} \biggr) 
-{2\over 9} \biggr] \,\,, \qquad  K = 6.663 \,\, . 
\end{equation} 
We remark again that the behavior near threshold,  
\begin{equation} 
{\rm Im}\, G_A(t) \sim (\sqrt t -3 M_\pi)^2 
\label{imgat}
\end{equation} 
is a stronger rise than  pure phase space 
\beq
\int\int_{z^2<1} d\omega_1 d\omega_2 \,
|\vec{l}_1| \, |\vec{l}_2| \sim (\sqrt t-3M_\pi)^3 \,\,\, .
\label{gaps}
\eeq
If one considers individual diagrams one finds that only the graphs (a), (b)
and (c) which have intermediate nucleon propagators show the stronger threshold
behavior of eq.(\ref{imgat}).
Diagram (d) without a nucleon propagator follows the phase
space argument eq.(\ref{gaps}). The obvious reason for these features is 
the coincidence of normal and
anomalous threshold in graphs (a), (b) and (c) which does not occur in diagram
(d) in the limit $m\to \infty$  
Finally, fig.~7 shows the isovector axial spectral function
Im\,$G_A(t)$ which again has a smooth rise from threshold and is about one
order of magnitude smaller than the isovector electric spectral function. Even
though Im\,$G_A(t)$ starts at order $q^5$ it is only three times as large as
the  isoscalar magnetic spectral function Im\,$G_M^S(t)$ of order
$q^7$.  We remark that  Im~$G_A(t)/t^3$, which has the 
same asymptotics as Im~$G^V_E(t)/t^2$ (to order $q^3$), is still monotonically
increasing from  threshold to $t=30 \, M_\pi^2$, indicating once more that the
three-pion continuum has no pronounced effect on the axial spectral
function near threshold. Of course, one also has to keep in mind that 
the available data for $G_A(t)$ are not yet that precise to perform 
accurate dispersion--theoretical calculation.

This completes our analysis of the spectral functions related to the isoscalar
electromagnetic and isovector axial nucleon form factors near threshold. We do
not find a visible strong enhancement slightly above threshold
$t_0=9\,M_\pi^2$ as in the isovector electromagnetic case, even though these
imaginary parts do not follow the phase space in the heavy nucleon mass
limit.   The presence of a nearby anomalous threshold at $t_c =
8.9\,M_\pi^2$ has no substantial effect on the strength of the
three--pion continuum.

%%%%%%%%%%%%%%%%%%%%%%%%%%%

\bigskip\bigskip

%%\newpage

\centerline{\Large {\bf Figure captions}}

\medskip

\begin{enumerate}
\item[Fig.1] Two--pion cut contribution to the nucleon isovector 
electromagnetic form factors. Wiggly, dashed and solid lines represent photons,
pions and nucleons, in order.

\item[Fig.2] Spectral distribution of the isovector electric and magnetic
nucleon form factors weighted with $1/t^2$ on the left wing of the
$\rho$--resonance as calculated in \cite{MMD}. Shown are Im\,$G_M^V (t)/t^2$
(upper solid line) and Im\,$G_E^V (t)/t^2$ (lower solid line). 
Suppressing the $\rho$--meson contribution as described in the text
$(F_\pi^V (t)\equiv 1)$ leads to the respective dashed--dotted lines.

\item[Fig.3] Spectral distribution of the isovector electric and magnetic
nucleon form factors weighted with $1/t^2$ calculated in heavy baryon CHPT.
Shown are Im\,$G_M^V (t)/t^2$ (upper lines) and Im\,$G_E^V (t)/t^2$ (lower 
lines). The solid and dashed lines refer to the order $q^4$ and $q^3$
calculations, respectively.

\item[Fig.4] Three--pion cut contribution to the nucleon isoscalar
electromagnetic form factors. The vertex denoted 'A' describes the
$\gamma\to 3\pi$ coupling and the one denoted 'B' the $3\pi \to \bar N N$
transition. 

\item[Fig.5] Two--loop diagrams contributing to the imaginary parts
of the isoscalar electromagnetic and isovector axial nucleon form factors. 

\item[Fig.6] Spectral distribution of the isoscalar electric and magnetic
nucleon form factors weighted with $1/t^2$ in the heavy nucleon limit.
Shown are Im\,$G_M^S (t)/t^2$ (upper line) and Im\,$G_E^S (t)/t^2$ (lower
line). 

\item[Fig.7] Spectral distribution of the  isovector axial nucleon form
factor weighted with $1/t^2$ in heavy nucleon limit.  

\end{enumerate}
%\end{document}

%\baselineskip 8pt plus 1pt minus 1pt

\newpage

%%%%%%%%%%%%%%%%%%  Fig. 1  %%%%%%%%%%%%%%%%%%%%%%%%%%%%%
\begin{figure}[bht]

%\vskip 5cm
$\;$\vspace{2cm}

\centerline{
\epsfysize=2in
\epsffile{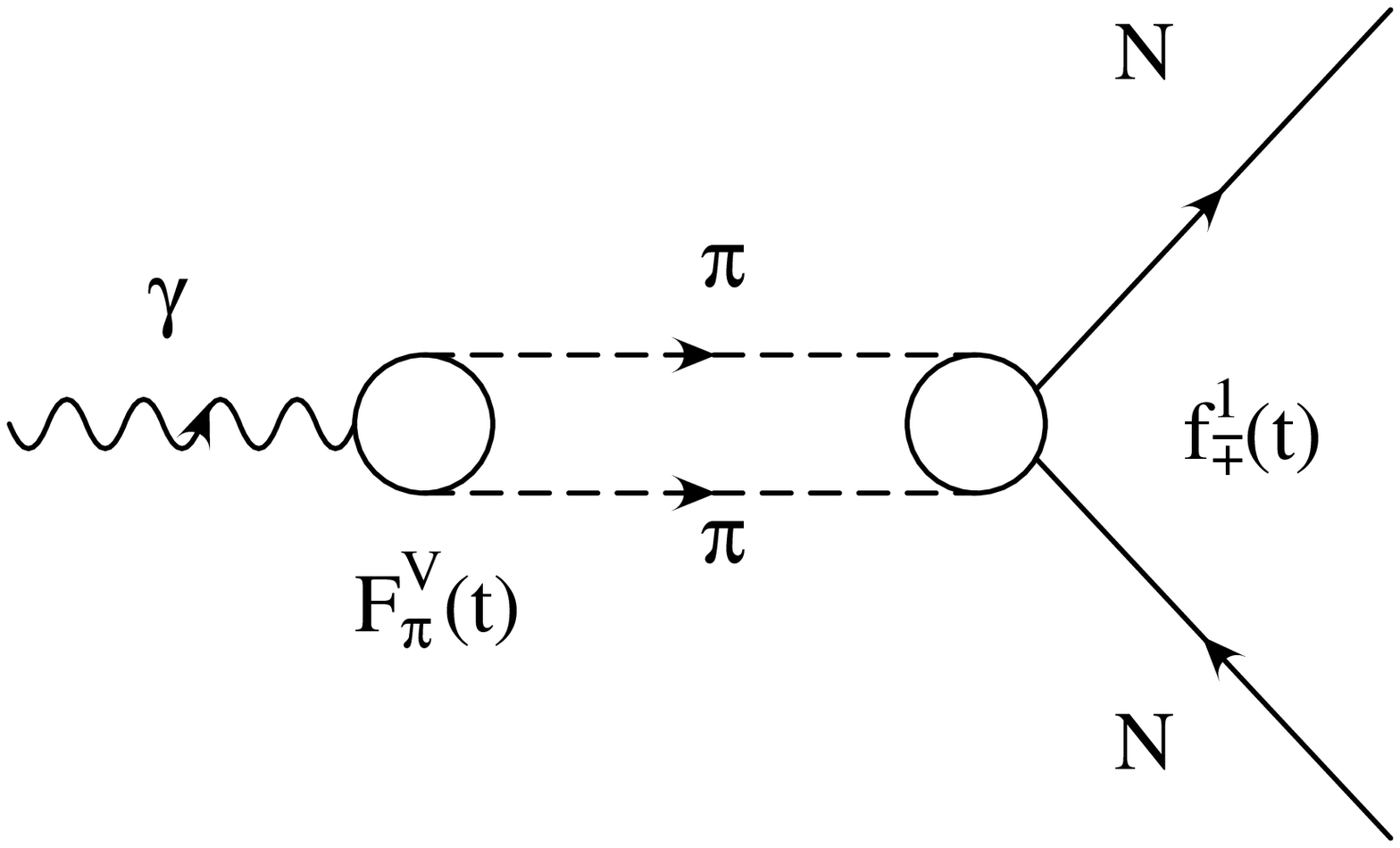}
}
%\bigskip\bigskip\bigskip\bigskip
\vskip 0.7cm

\centerline{\Large Figure 1}
%\caption[]{}
\end{figure}
 
%%%%%%%%%%%%%%%%% END FIGURE  %%%%%%%%%%%%%%%%
%%%%%%%%%%%%%%%%%%%%%%%%%%%  Fig. 2  %%%%%%%%%%%%%%%%%%%%%%%%%%%%%
%\newpage
$\;$\vspace{0.3cm}

\begin{figure}[bht]
\centerline{
\epsfysize=4in
\epsffile{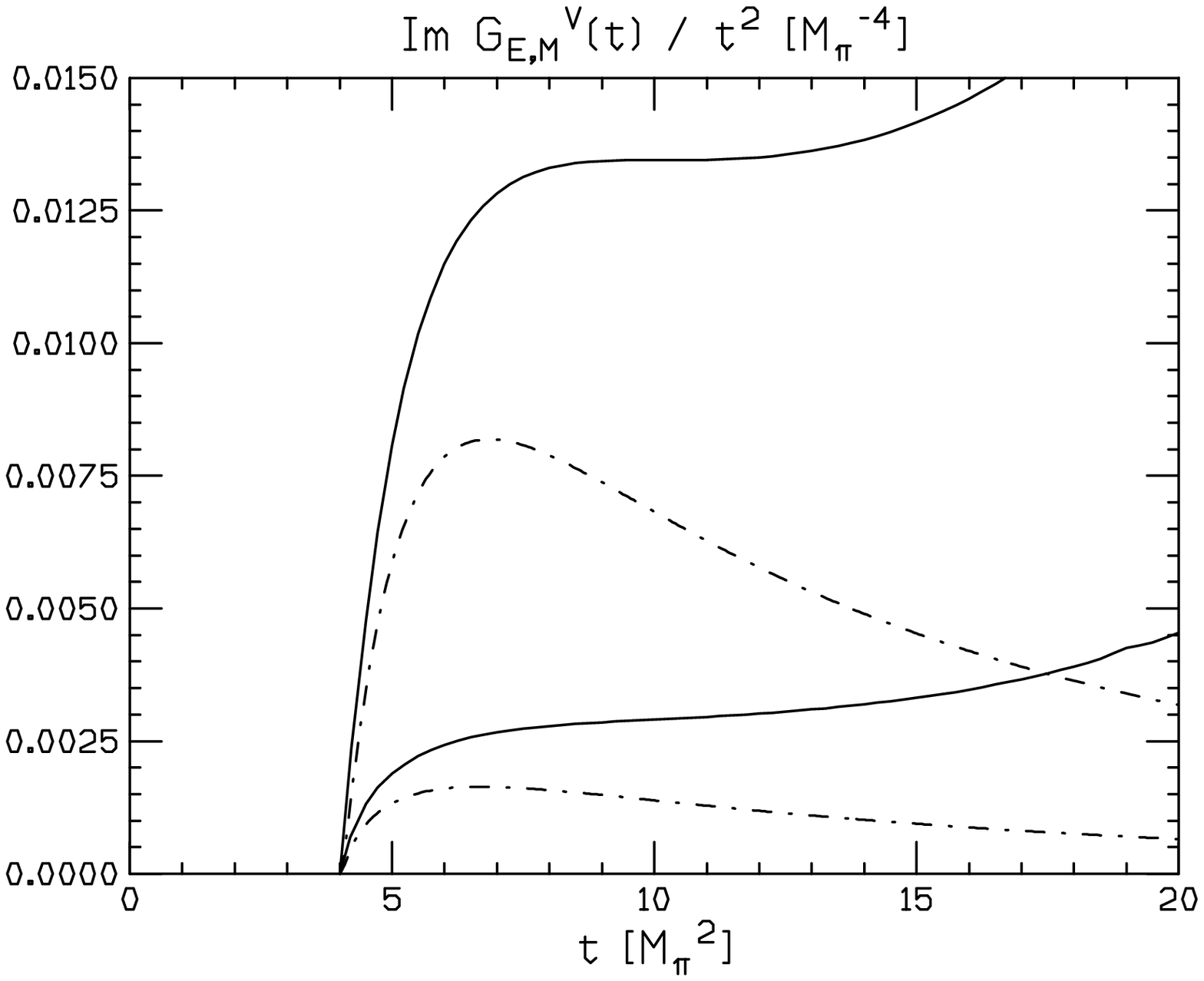}
}
%\bigskip\bigskip\bigskip\bigskip
\vskip 0.7cm

\centerline{\Large Figure 2}
%\caption[]{}
\end{figure}
 
%%%%%%%%%%%%%%%%% END FIGURE  %%%%%%%%%%%%%%%%%%%%%%%%%

\newpage

%%%%%%%%%%%%%%%%%%%%%%%%%%%  Fig. 3  %%%%%%%%%%%%%%%%%%%%%%%%%%%%%
%\newpage
$\;$\vspace{0.3cm}

\begin{figure}[bht]
\centerline{
\epsfysize=4in
\epsffile{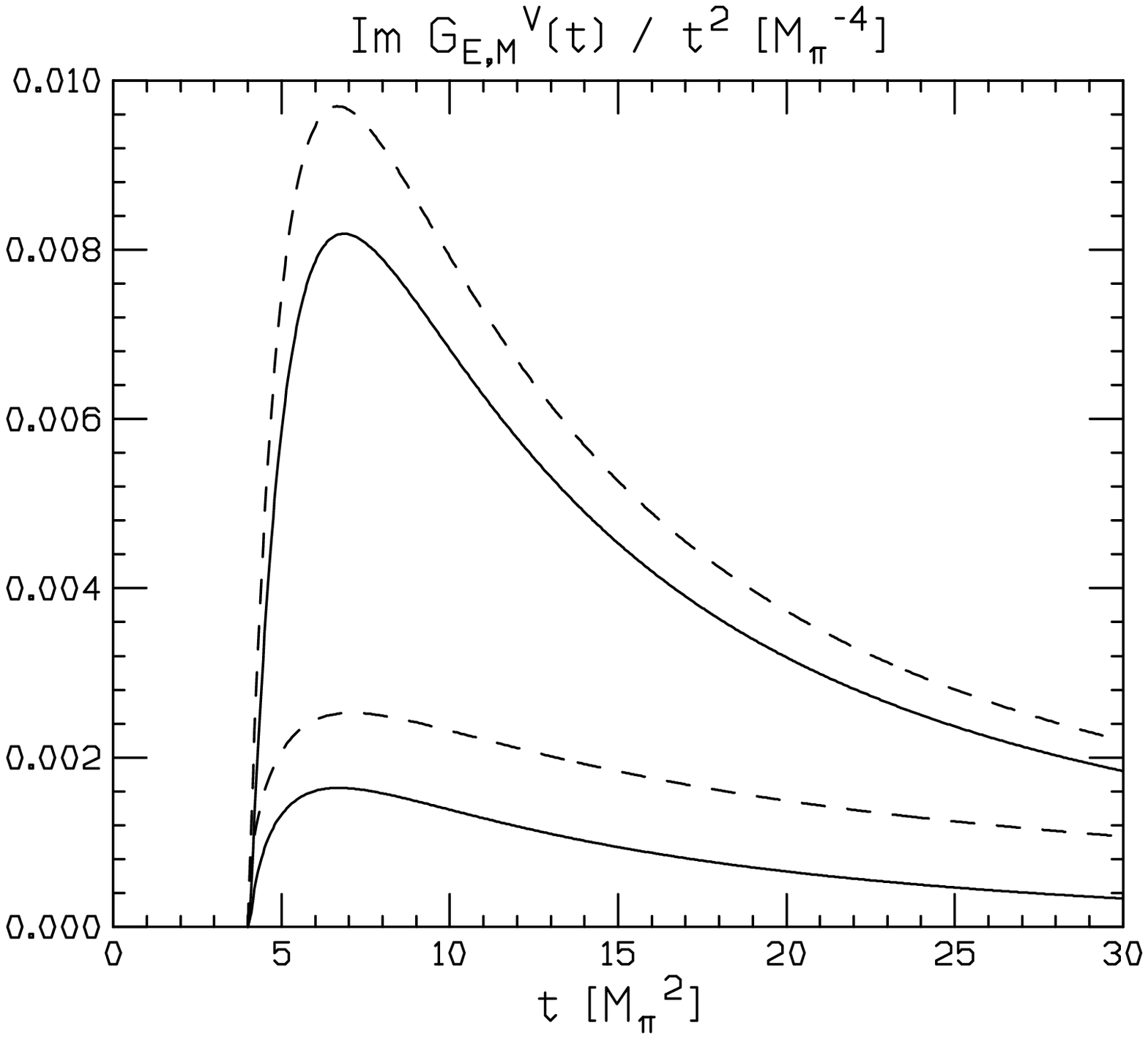}
}
%\bigskip\bigskip\bigskip\bigskip
\vskip 0.7cm

\centerline{\Large Figure 3}
%\caption[]{}
\end{figure}
 
%%%%%%%%%%%%%%%%% END FIGURE  %%%%%%%%%%%%%%%%%%%%%%%%%

%%%%%%%%%%%%%%%%%%  Fig. 4  %%%%%%%%%%%%%%%%%%%%%%%%%%%%%
\begin{figure}[bht]

%\vskip 5cm
$\;$\vspace{2cm}

\centerline{
\epsfysize=2in
\epsffile{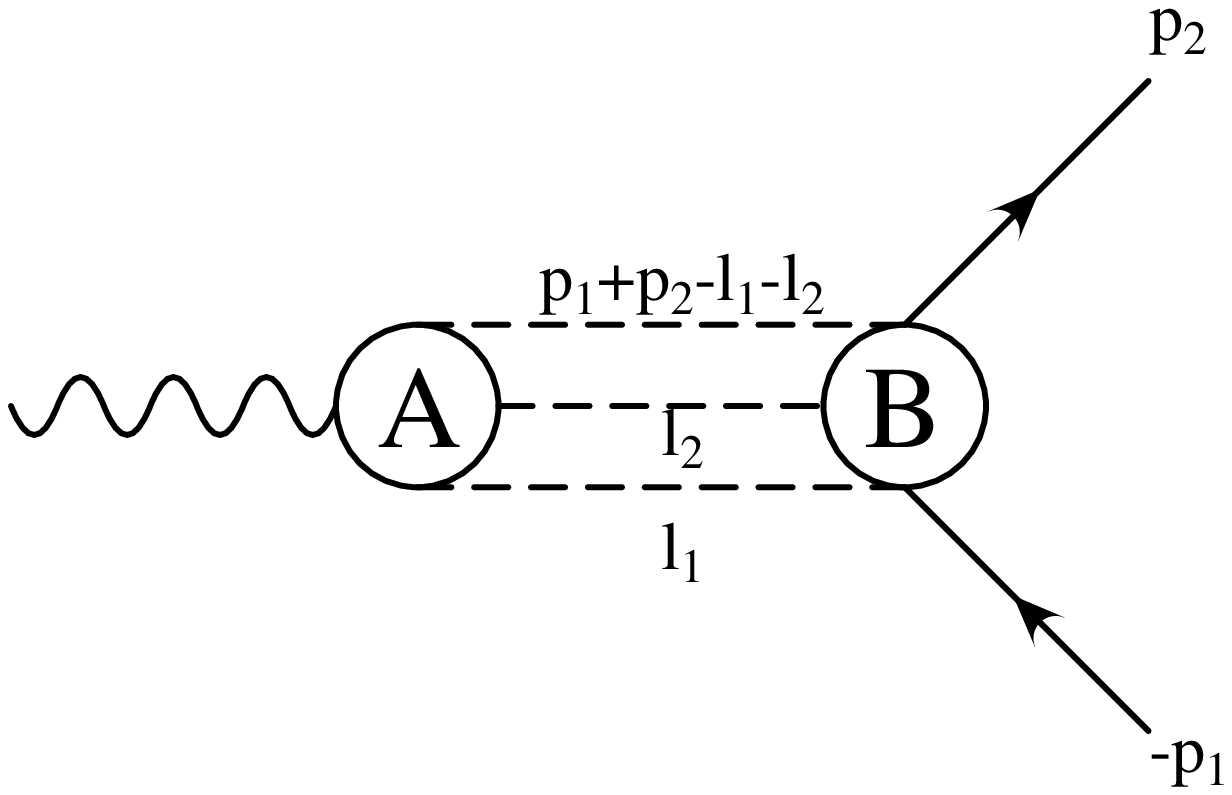}
}
%\bigskip\bigskip\bigskip\bigskip
\vskip 0.7cm

\centerline{\Large Figure 4}
%\caption[]{}
\end{figure}
 
%%%%%%%%%%%%%%%%% END FIGURE  %%%%%%%%%%%%%%%%

\newpage

%%%%%%%%%%%%%%%%%%%%%%%%%%%  Fig. 5  %%%%%%%%%%%%%%%%%%%%%%%%%%%%%
%\newpage
$\;$\vspace{0.3cm}

\begin{figure}[bht]
\centerline{
\epsfysize=3.5in
\epsffile{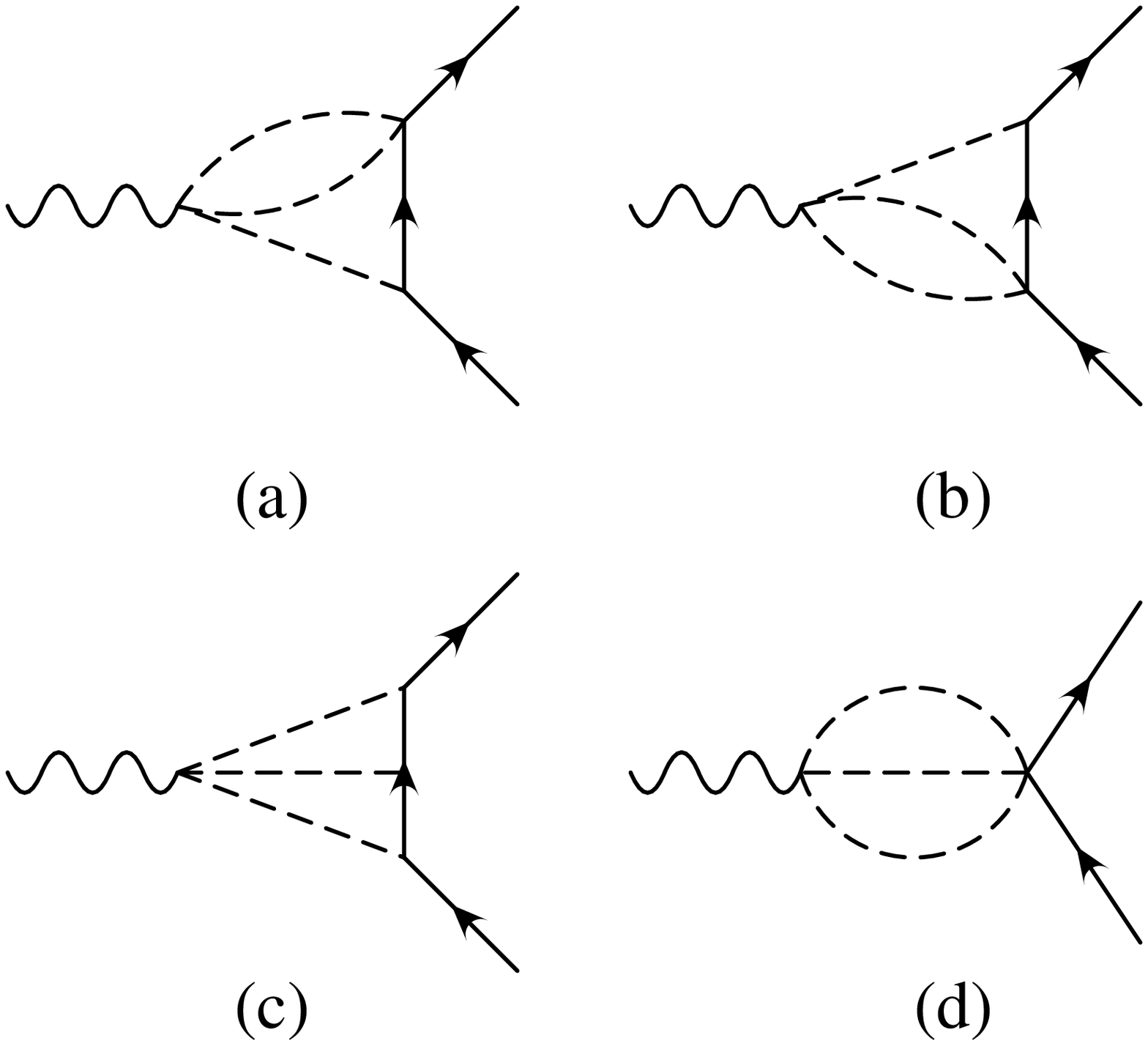}
}
%\bigskip\bigskip\bigskip\bigskip
\vskip 0.7cm

\centerline{\Large Figure 5}
%\caption[]{}
\end{figure}
 
%%%%%%%%%%%%%%%%% END FIGURE  %%%%%%%%%%%%%%%%%%%%%%%%%

%%%%%%%%%%%%%%%%%%%%%%%%%%%  Fig. 6  %%%%%%%%%%%%%%%%%%%%%%%%%%%%%
%\newpage
$\;$\vspace{0.3cm}

\begin{figure}[bht]
\centerline{
\epsfysize=3in
\epsffile{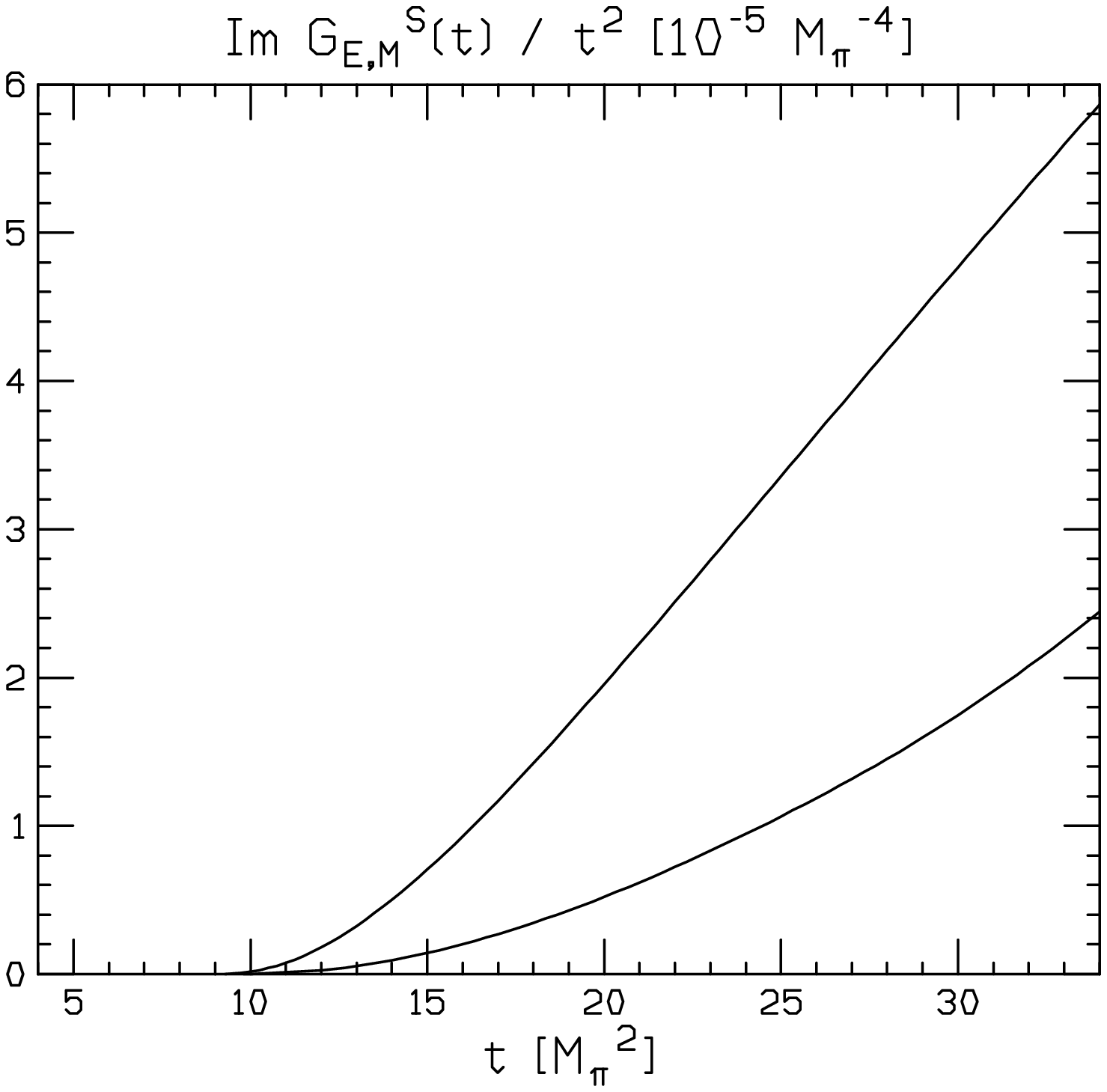}
}
%\bigskip\bigskip\bigskip\bigskip
\vskip 0.7cm

\centerline{\Large Figure 6}
%\caption[]{}
\end{figure}
 
%%%%%%%%%%%%%%%%% END FIGURE  %%%%%%%%%%%%%%%%%%%%%%%%%

\newpage

%%%%%%%%%%%%%%%%%%%%%%%%%%%  Fig. 7  %%%%%%%%%%%%%%%%%%%%%%%%%%%%%
%\newpage
$\;$\vspace{3cm}

\begin{figure}[bht]
\centerline{
\epsfysize=4in
\epsffile{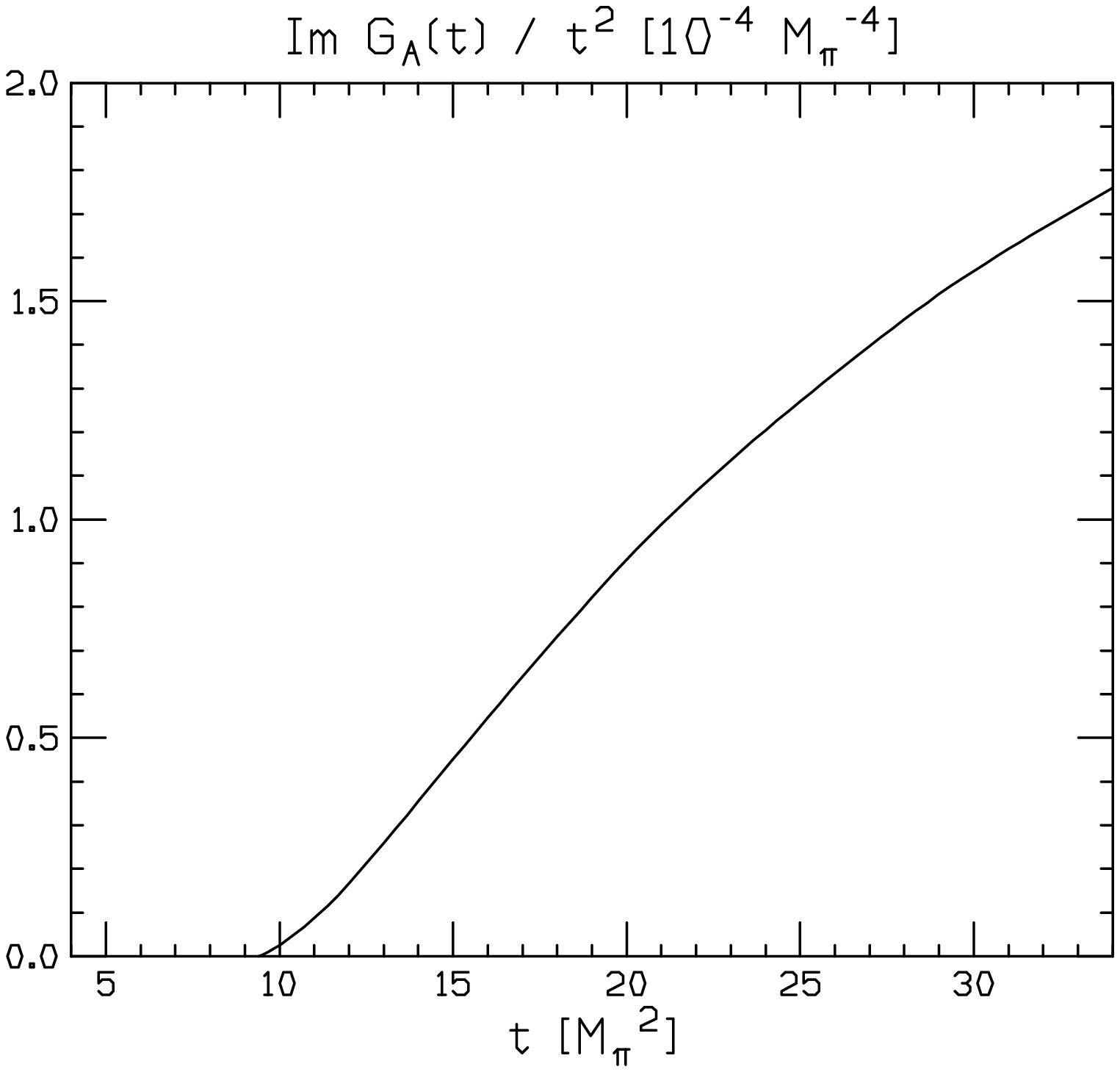}
}
%\bigskip\bigskip\bigskip\bigskip
\vskip 0.7cm

\centerline{\Large Figure 7}
%\caption[]{}
\end{figure}
 
%%%%%%%%%%%%%%%%% END FIGURE  %%%%%%%%%%%%%%%%%%%%%%%%%

\end{document}